\documentstyle[12pt]{article}
\input epsf
\newwrite\ffile\global\newcount\figno \global\figno=1

\def\writedef#1{}
\def\figin{\epsfcheck\figin}\def\figins{\epsfcheck\figins}
\def\epsfcheck{\ifx\epsfbox\UnDeFiNeD
\message{(NO epsf.tex, FIGURES WILL BE IGNORED)}
\gdef\figin##1{\vskip2in}\gdef\figins##1{\hskip.5in}
\else\message{(FIGURES WILL BE INCLUDED)}%
\gdef\figin##1{##1}\gdef\figins##1{##1}\fi}
\def\figinsert{}
\def\ifig#1#2#3{\xdef#1{fig.~\the\figno}
\writedef{#1\leftbracket fig.\noexpand~\the\figno}%
\figinsert\figin{\centerline{#3}}\medskip\centerline{\vbox{\baselineskip12pt
\advance\hsize by -1truein\center\footnotesize{  Fig.~\the\figno.} #2}}
\bigskip\endinsert\global\advance\figno by1}
\def\endinsert{}

\begin{document}

\begin{flushright}
RU--98--13\\
\today 
\end{flushright}

\begin{center}
\bigskip\bigskip
{\Large \bf  The Spectrum of Softly Broken  $N=1$ Supersymmetric
Yang-Mills Theory}
\vspace{0.3in}      

{\bf  G.~R. ~Farrar,~ G. Gabadadze~and ~M. Schwetz}
\vspace{0.2in}

{\baselineskip=14pt
Department of Physics and Astronomy, Rutgers University \\
Piscataway, New Jersey 08855, USA}\\
{\rm emails: farrar, gabad, myckola@physics.rutgers.edu}

\vspace{0.2in}
\end{center}

\vspace{0.9cm}
\begin{center}
{\bf Abstract}
\end{center} 
\vspace{0.3in}
We study the spectrum of the softly broken {\it generalized}
Veneziano-Yankielowicz
effective action for $N=1$ SUSY Yang-Mills theory. 
Two dual formulations of the effective action are  given. 
The spurion method is used for the soft SUSY breaking.
Masses of the bound states are calculated and 
mixing patterns are discussed. Mass splittings of pure gluonic states 
are consistent with predictions of conventional Yang-Mills theory. 
The  results can be tested in lattice simulations of the 
SUSY Yang-Mills model. 
\vspace{2cm}


\newpage
\noindent {\bf Introduction}
\vspace{0.2in}

Some time ago great progress was made in understanding
the ground state structure of many supersymmetric gauge theories \cite {S},
\cite {SW}. It is highly desirable to have  
direct non-perturbative tests of
those results. There is a possibility that these models can  be simulated 
on the lattice. Some preliminary work 
toward this complicated task has already been
performed (see refs. \cite{Lattice}, \cite{lattice:cernrome}).

The lattice regularization violates 
supersymmetry \cite{CV}. Thus, some special fine-tuning is required to
recover the SUSY limit on the  lattice.
Away from the SUSY point, the continuum
limit of the lattice theory is described by a model with explicit SUSY  
breaking  terms. In some cases  those terms  may trigger only
soft SUSY  breaking \cite{giradello}, 
although this is not guaranteed in general.
 
Softly broken SUSY models can be studied using the spurion technique
\cite {spurion}.
Some ``exact'' results were obtained within that  approach
\cite{soft1,soft2,soft3}.
In this paper we consider softly broken   supersymmetric 
Yang-Mills theory, the model which is relevant for lattice simulations. 
At the classical level supersymmetric Yang-Mills (SYM) is 
a theory with only one parameter,
the gauge coupling constant. The lowest-dimensional renormalizable 
SUSY breaking  term  allowed by gauge invariance is the gaugino mass term.
Therefore, we consider SYM with a  gaugino mass term as a theory
describing the continuum limit of the lattice regularized action.

In analogy with QCD, one expects that the spectrum
of this  model consists of colorless bound states of gluinos and gluons.
Among those are: pure gluonic bound states (glueballs), 
gluino-gluino mesons  and gluon-gluino composites. 
These states fall into the lowest-spin representations 
of the $N=1$ SUSY algebra written in the basis of parity eigenstates
\cite {us}. The masses and interactions of these bound states can  be 
given within the effective Lagrangian approach. 
The effective action for $N=1$ SYM was 
proposed by Veneziano and Yankielowicz (VY) \cite {VY}. 
The VY action \cite {VY} involves fields for gluino-gluino 
and gluino-gluon bound states.  However, it does not include dynamical
degrees of freedom which would  correspond to pure gluonic 
composites (glueballs).

We argued in ref. \cite {us} that there are  no physical reasons  to 
expect glueballs to be heavier and decoupled from gluino-gluino
and gluino-gluon bound states in $N=1$ SUSY YM theory.
Moreover, there are SQCD sum rule based arguments indicating that 
the low-energy spectrum of SYM theory is not exhausted by
the gluino containing bound states only \cite {Shap};
glueball degrees of freedom should also be taken into account.  

The generalization of the VY effective action that includes
pure gluonic degrees of freedom was given in ref. \cite {us}.
The generalized VY effective Lagrangian of ref. \cite {us}
describes mixed states of glueballs, gluino-gluino and gluino-gluon 
bound states. 
The fundamental superfield upon which that  construction 
of the generalized VY action is based \cite {us}
is a constrained tensor superfield \cite {Gates}. The set of
components of that superfield 
includes as a subset the VY chiral supermultiplet. 

The aim of the present paper is twofold.
First we propose a  new representation of the generalized VY effective 
action of ref. \cite {us}. This action  is equivalent 
to the previously proposed  one \cite {us}, but  it  uses 
two different chiral supermultiplets instead of the 
tensor supermultiplet approach adopted in \cite {us}. 

Then we introduce soft SUSY breaking terms 
in the generalized VY Lagrangian and study mass splittings  and mixing
patterns in the softly broken theory. These results can be directly 
tested in lattice calculations. Predictions  
for the masses of the gluino-gluino and gluon-gluino bound states  
and their splittings in the broken theory  were made  in
ref. \cite{NSM} using  the original VY effective
action \cite{VY} and the spurion technique. 
We will see that the presence of the glueball degrees
of freedom  changes the vacuum state of the broken theory. 
As a result, the mass splittings are also modified. 

The paper is organized as follows. In section 1 we briefly review 
the generalized VY effective Lagrangian and recall  some results 
obtained in ref. \cite {us}. In section 2 we explain how one can
reformulate the generalized VY Lagrangian in terms of two independent 
chiral superfields using the chiral-tensor superfield  duality \cite
{Gates}, \cite {GGRZ}.
In section 3 we show how the effective action is modified when 
the gaugino mass term is introduced in SYM  through the 
spurion method. Section 4 reports  the masses and mixings for physical
eigenstates of the broken theory.
\vspace{0.3in}\\
{\bf 1. The Generalized VY Effective Action }
\vspace{0.2in}

The on-shell Lagrangian of SYM for an  $SU(N_c)$ gauge group  
is
\begin{eqnarray}
{\cal L} ~=~ \frac{1}{g^2}\left[ \,-\frac{1}{4} 
G_{\mu\nu}^a G_{\mu\nu}^a
~+~ i\lambda_{\dot\alpha}^\dagger D^{\dot\alpha\beta}\lambda_\beta
\right] ~. 
\nonumber
\end{eqnarray}
In terms of superfields the expression above can be written
\begin{eqnarray}
{\cal L} ~=~ \int d^2 \theta~
\frac{1}{8\pi} Im\, \tau  W^{\alpha}W_{\alpha}~+h.c., 
\label{lagrangian}
\end{eqnarray}
where the gauge coupling is defined to be
$\tau = \frac{4\pi i}{g^2} + \frac{\theta_0}{2\pi}$. 
For the purposes of this paper we set the theta term to be equal to
zero,  $\theta_0 =0$.

The classical action of  $N=1$ SYM theory is invariant  
under $U(1)_R$, scale 
and superconformal transformations. 
In the 
quantum theory these symmetries are 
broken by the  chiral, scale and superconformal  anomalies respectively.
Composite operators that appear  in the expressions for the 
anomalies can be thought of as  component fields  
of a chiral supermultiplet $S$ \cite {WessZumino}
\begin{eqnarray}
S\equiv { \beta (g) \over 2 g} W^{\alpha}W_{\alpha}\equiv
A(y)+\sqrt{2} \theta \Psi(y) + \theta^2 F(y),  \nonumber
\end{eqnarray}
where $\beta(g)$ is  the SYM beta function for which the 
exact expression is  known  \cite {beta}. The lowest component
of the $S$ supefield is bilinear in gluino fields and has the quantum
numbers of the scalar and pseudoscalar gluino bound states.  The
fermionic component in $S$ describes the gluino-gluon composite and
the $F$ component of the chiral superfield includes operators
corresponding to both the scalar and 
pseudoscalar glueballs ($G_{\mu\nu}^2$ and 
$G_{\mu\nu}{\tilde G^{\mu\nu}}$ respectively) \cite {VY}. 

Assuming that the effective action (more precisely,  
the generating functional for one-particle-irreducible (1PI) 
Green's functions \cite {GoldstoneSalamWeinberg}) of the 
model can be written in terms of the single  superfield $S$,   and 
requiring also that the effective action respects all the global
continuous symmetries and 
reproduces the anomalies
of the SYM theory,   one derives the Veneziano-Yankielowicz 
effective action \cite {VY}. Let us mention that
the actual variables, in terms of which the  generating functional for 
the 1PI Green's functions (or effective action in our conventions)
is written, are the VEV's of composite operators calculated at nonzero
values of external sources \cite {Shore}. In this paper, as well as
in ref.  \cite {us},  we use a simpified notation where 
the VEV's are denoted by the corresponding composite operators. 

It was noticed in ref. \cite
{ShifmanKovner} that the VY action does not respect the discrete
$Z_{2N_c}$ symmetry -- the nonanomalous remnant of anomalous 
$U(1)_R$ transformations.
The VY action was amended by an appropriate term which
makes the action invariant under 
the discrete $Z_{2N_c}$ group \cite {ShifmanKovner}\footnote[9]{The vacua
with the broken chiral symmetry  are labeled by an  integer
$n=0,..., N_c-1$. In this work we study the spectrum of the model about the 
$n=0$ ground state.}. 

However, as we mentioned above, the VY action does not include
all possible lowest-spin bound states of  SYM theory.
Glueballs  are missing in that description because they are only
present in the auxilliary component of the $S$ superfield and can be
integrated out.  In ref. \cite {us}, in order to 
account for glueball degrees of freedom, we proposed 
to formulate the effective action in terms of a more general superfield,
the real tensor superfield $U$ \cite {Gates}. The superfield $U$ 
can be written in component form as follows: 
\begin{eqnarray}
U=B+i\theta \chi -i {\bar \theta} {\bar \chi}+{1\over 16}\theta^2 {A^*}+
{1\over 16} {\bar \theta}^2 A+{1\over 48 }\theta \sigma^\mu {\bar
\theta} \varepsilon_{\mu\nu\alpha\beta}C^{\nu\alpha\beta}+ 
\nonumber \\
{1\over 2} \theta^2 {\bar \theta} \left ( {\sqrt{2} \over 8}{\bar
\Psi} +{\bar \sigma}^\mu \partial_\mu \chi \right )+
{1\over 2}{\bar  \theta}^2 \theta  \left ( {\sqrt{2} \over 8}
\Psi - \sigma ^\mu \partial_\mu {\bar \chi }\right )+{1\over 4}
\theta^2 {\bar \theta^2} \left ( {1\over 4} \Sigma -\partial^2 B\right ).
\label{U}
\end{eqnarray} 
It is straightforward   to show that the 
real superfield $U$  satisfies the relation\footnote{Despite a 
seeming similarity, the tensor multiplet $U$ should not be
interpreted as a usual vector multiplet. The vector field which might
be introduced in this approach as a Hodge dual of  the three-form 
potential $C_{\mu\nu\alpha}$ would give
mass terms with the wrong sign in our approach (see ref. \cite {us}),
thus,  the actual physical variable  is the three-form potential  
$C_{\mu\nu\alpha}$ rather than its dual vector field 
(the Chern-Simons current).} 
\begin{eqnarray}
S=-4 {\bar D}^2 U,
\nonumber
\end{eqnarray}  
where the $F$ term of the chiral supermultiplet $S$
is related to the fields $\Sigma$ and $C_{\mu\nu\alpha}$
in the following way\footnote[3]{In this  notation
$\Sigma$ is proportional to $G_{\mu\nu}^2$ and 
$\varepsilon_{\mu\nu\alpha\beta}\partial^{\mu}
C^{\nu\alpha\beta}$ is proportional to $G_{\mu\nu}{\tilde
G^{\mu\nu}}$ \cite{us}.}
\begin{eqnarray}
F=\Sigma+i{1\over 6}\varepsilon_{\mu\nu\alpha\beta}\partial^{\mu}
C^{\nu\alpha\beta}, \nonumber
\end{eqnarray} 
and $A$ and $\Psi$ are respectively the scalar and fermion
components of the superfield $S$. 

We argued \cite {us} that 
the effective Lagrangian  for the lowest-spin multiplets 
of the $N=1$ SYM theory can be written in terms of the 
$U$ field only.  That Lagrangian takes the following 
form \cite {us}
\begin{eqnarray}
{\cal L}={1\over \alpha} (S^{+}S)^{1/3}\Big|_D+ 
\gamma \Big [( S \log {S\over \mu^3}-S)\Big|_F+{\rm h.c.}\Big ]+   
{1\over \delta} \left ( -{U^2\over (S^+S)^{1/3}} \right )\Big|_D,
\label{NewA}
\end{eqnarray}
where $\alpha~{\rm and}~\delta$ are arbitrary positive constants
and $\gamma =- (N_c g/16 \pi^2 \beta(g))>0$. 
Notice that the superfield $S$ is not an 
inpendent variable in this Lagrangian.
It is rather related to the $U$ superfield through the formula
\begin{eqnarray}
S-\langle S \rangle= -4 {\bar D}^2 U. 
\nonumber
\end{eqnarray}
In the above equation we took into account that 
the $S$ superfield has a nonzero VEV in the 
phase where chiral symmetry is broken,  
$\langle S \rangle \equiv \mu^3$.
Thus, the only independent superfield in the Lagrangian 
(\ref {NewA}) is the $U$ field.  

In this approach the following fields become dynamical \cite {us}:
\begin{itemize}

\item The $B$ field propagates and it represents
one massive real scalar degree of freedom
(identified with the scalar glueball).

\item The three-form potential $C_{\mu\nu\alpha}$, which becomes massive,
also propagates. It represents one physical  degree of freedom 
(identified with the pseudoscalar glueball). 

\item The complex field $A$,  being 
decomposed into parity eigenstates,   describes the massive gluino-gluino 
scalar and pseudoscalar mesons.

\item $\chi$ and $\Psi$ describe
the massive gluino-gluon fermionic bound states. 

\end{itemize}
Studying the potential of the model, we found  that
the physical eigenstates fall into two different 
mass ``multiplets'' (see  ref. \cite {us} for details).  
Neither of them contain pure gluino-gluino, gluino-gluon or
gluon-gluon bound states.  Instead, the physical excitations are 
mixed states of these composites.  The heavier set of states contains: 
\begin{itemize}
\item A pseudoscalar meson, which without mixing reduces to the $0^{-+}$
gluino-gluino bound state (the analog of the QCD $\eta'$ meson).
\item A scalar meson that without mixing is a $0^{++}$ $(l=1)$ 
gluino-gluino excitation.
\item A fermionic gluino-gluon bound state. 
\end{itemize}
These heavier states form  the chiral supermultiplet described by the VY
action. That action  is recovered in the 
$\delta \rightarrow \infty$ limit.  
The new states which appear as a result of our 
generalization form  the lighter multiplet:
\begin{itemize}
\item A scalar meson, which without  mixing is  a $0^{++}$ ($l=0$)
glueball. 
\item A pseudoscalar state, which for zero  mixing is identified as a
$0^{-+}$ ($l=1$) glueball.
\item A  fermionic gluino-gluon bound state. 
\end{itemize}

Notice, that although the physical states fall into
multiplets whose $J^P$ quantum numbers correspond to two chiral 
supermultiplets,
the action was written in terms of the one real tensor supermultiplet $U$.
In particular, the pseudoscalar glueball in this approach is
described by the only physical component of the massive three-form potential
$C_{\mu\nu\alpha}$. The field strength of that potential couples to the
pseudoscalar gluino-gluino bound state as it would couple  to the $\eta'$
meson in QCD \cite{veneziano}\footnote[5]
{The three-form potential proved to be
useful  for the description of the pseudoscalar glueball in
conventional Yang-Mills theory \cite {gabad}.}. 

Since the physical spectrum of the mixed states fall into multiplets 
whose spin-parity  quantum numbers correspond to two chiral 
supermultiplets, one might  be wondering about the possibility to  
rewrite the whole action it terms of two different chiral superfields. 
If that is possible it would be crucial to study
what peculiarities of the two-chiral-multiplet action
allow  it to be written in terms of only a  real 
supermultiplet $U$,
as was done  in ref. \cite {us}. In the next section  
we address these questions. 
\vspace{0.3in} \\
{\bf 2. The Two Chiral Supermultiplet Action}
\vspace{0.2in}

The relation between a real tensor and chiral
supermultiplets  (the so called chiral-linear duality) 
was established in ref. \cite {Gates}.
For SYM theory the chiral-linear duality was used in ref. \cite 
{Derendinger} (see also discussions in refs. \cite {Binetruy}). 
Applied to our problem the results  of refs. \cite
{Gates}, \cite {Derendinger} and \cite {Binetruy}
can be stated as follows. One introduces into the effective 
Lagrangian a new chiral superfield, let us denote it by  $\chi$
\begin{eqnarray}
\chi (y, \theta) \equiv \phi_{\chi}(y)+ \sqrt {2}\theta
\Psi_{\chi}(y)+\theta^2 F_{\chi}(y).
\label{chi} 
\end{eqnarray}
One can find an  effective Lagrangian written in terms
of two chiral superfields,  $S$ and $\chi$  which is 
equivalent to the expression given in (\ref {NewA}). 
In our case 
\begin{eqnarray}
{\cal L}~=~{1\over \alpha}\, (S^{+}\,S)^{1/3} \Big|_D ~+~ 
{\delta \over 4 }\, (S^{+}\,S)^{1/3}\, (\chi~+~ \chi^{+})^2 \Big|_D
~+~ \nonumber \\
\Big[\gamma \,( S \,\log {S\over \mu^3}~-~S)\Big|_F ~+~ 
{1 \over 16} \,\chi\, (S ~-~ \mu^3)\Big|_F~+~{\rm h.c.}\Big]~.
\label{2s}
\end{eqnarray}
Comparing this expression to the VY Lagrangian one 
notices that both the K{\"a}hler potential and the superpotential
are modified by new terms. The  multiplets $S$ and $\chi$ are independent.

We would like to relate  this expression to 
the Lagrangian of the theory written in terms of the $U$ field (\ref {NewA}). 
If the $U$ field is postulated  as a fundamental degree of freedom,
then the $S$ field is a  derivative superfield
\begin{eqnarray}
S= \mu^3-4 {\bar D}^2 U.  
\label{USrelation}
\end{eqnarray}
Using this relation 
the Lagrangian (\ref {2s}) can be rewritten as 
\begin{eqnarray}
{\cal L}~=~{1\over \alpha}\, (S^{+}\,S)^{1/3} \Big|_D ~+~ 
{\delta \over 4 }\, (S^{+}\,S)^{1/3}\, (\chi~+~ \chi^{+})^2 \Big|_D
~+~ \nonumber \\
\Big[\gamma \,( S \,\log {S\over \mu^3}~-~S)\Big|_F ~+~{\rm
h.c.}\Big]~+~ U (\chi~+~ \chi^{+})\Big|_D~.
\label{2cs}
\end{eqnarray}
This expression depends on two superfields $U$ and $\chi$ ($S$ 
is expressed through $U$ in accordance with (\ref {USrelation})). However, 
the dependence on the chiral superfield $\chi$ is trivial, 
the combination  $\chi +\chi^{+}$ can be integrated out 
from the Lagrangian  (\ref {2cs}). As a result one derives 
\begin{eqnarray}
\chi + \chi^{+}= -{ 2U\over \delta (S^{+}S)^{1/3}}.
\label{Uchi}
\end{eqnarray}
Substituting this expression back into the Lagrangian
(\ref {2cs}) one arrives at  the original  expression (\ref {NewA})
where the $S$ field is a derivative field satisfying 
the relation (\ref {USrelation}).

Let us stress again that the descriptions in terms of the Lagrangian 
(\ref {NewA}) and (\ref {2s}) are equivalent on the mass-shell. 
In the Lagrangian (\ref {NewA}) the dynamical degrees of freedom are 
assigned to  the only superfield $U$, while in the Lagrangian  (\ref
{2s}) the physical degrees of freedom are found  as components of two
chiral  supermultiplets $S$ and $\chi$. The peculiarity of 
the expression (\ref {2cs}) is that the chiral superfield 
$\chi$ enters only through the real combination  $\chi+\chi^+$.
That is why it was possible  to formulate the action in terms only 
of the real superfield $U$.  It is essential from a physical point of
view since the component glueball field must be real.

Using the Lagrangian (\ref {2s}) one calculates  the potential of the 
supersymmetric model. Integrating out the auxiliary fields of 
both chiral multiplets one finds 
\begin{eqnarray}
V_0= {2\over \delta (16)^2} {|\phi^3-\mu^3|^2
\over |\phi |^2}~+~
{9\alpha |\phi |^4 \over 1- {\alpha\over \delta} 
{B^2\over |\phi |^4}}\cdot  \left |{\phi_{\chi}\over 16}
~+~3\gamma \log {\phi \over
\mu}~+~{B(\phi^3-\mu^3)\over 24 \delta |\phi|^2 \phi^3}\right |^2, 
\label{Potential}
\end{eqnarray}
where the following notations are adopted 
\begin{eqnarray}
\phi_{\chi}={1\over \sqrt {2}} (\sigma +i\pi),~~~~~~~~~
\sigma \equiv -{\sqrt {2} B\over \delta |\phi|^2}. \nonumber
\nonumber
\end{eqnarray}
The minimum of this potential is located at the point in  field
space where 
$\langle \phi  \rangle =\mu$, $\langle B \rangle =
\langle \phi_{\chi}\rangle=0 $. The potential (\ref {Potential})
is positive definite  for field configurations satisfying
$\alpha B^2 < \delta |\phi|^4$. Since the VEV of 
the $\phi$ field is nonzero and the VEV of the $B$ field is
zero  the positivity condition is satisfied for  small
oscillations about the SUSY minimum specified above. Notice that
all SUSY  field configurations are confined within a  valley with infinite 
potential walls encountered at $\alpha B^2=\delta |\phi|^4$. 
Thus, the potential (\ref {Potential}) and the Lagrangian 
(\ref {2s}) itself describe only small oscillations 
about the SUSY minimum. In general, some higher order
polynomials in the $\chi$ (or $U$) field could be present 
in the effective Lagrangian. In this work 
we are interested only in the mass spectrum 
of the model, so  the approximation we used above is good enough
for our goals.  

In the next
section we introduce soft SUSY breaking terms in the effective
Lagrangian and study minima  and the spectrum of the corresponding
potential.  
\vspace{0.3in} \\
{\bf 3. Soft SUSY Breaking }
\vspace{0.2in}

The gaugino mass term can  be introduced in the Lagrangian 
(\ref {lagrangian}) by means of 
the parameter $\tau$. One regards 
$\tau$ as a chiral superfield \cite {spurion}. A
nonzero VEV of the  $F$ component of $\tau$ yields a
SUSY breaking gaugino mass term in 
(\ref {lagrangian}). Thus, one performs 
the following substitution in expression (\ref {lagrangian}) 
$$\tau \rightarrow \tau + F_{\tau} \theta \theta.$$ 
As a result, the following new term appears 
in the Lagrangian  of SYM 
\begin{eqnarray}
- {1 \over 8 \pi}\, Im\, [\, F_{\tau}\, \lambda \lambda \,]+{\rm h.c.}.
\nonumber
\end{eqnarray}
To make the gaugino mass canonically normalized  one sets
$F_{\tau} = i\, 8 \pi\, m_\lambda /g^2$.
In the low-energy theory the $\tau$ parameter 
enters through the dynamically
generated scale of the theory
$\mu=\mu_0 {\rm exp}(-{8\pi^2\over \beta_0 g^2(\mu_0)})=
\mu_0 {\rm exp}(i{2\pi \tau \over \beta_0 })$. After  the 
$\tau$ parameter is claimed to be a chiral superfield one should
regard the $\mu$ parameter as a chiral superfield too. Thus,
one also makes the following substitution in the low-energy effective
Lagrangian of the model
\begin{eqnarray}
\mu \rightarrow \mu~{\rm exp}\left (-{16\pi^2 m_\lambda \over g^2 \beta_0}   
\theta \theta \right ),
\nonumber
\end{eqnarray}
where $\beta_0$ stands for the first coefficient of the beta function. 
Performing this redefinition of the $\mu$ parameter 
in the Lagrangian (\ref {2s})  one finds 
the following  additional term  in the scalar potential of the model 
\begin{eqnarray}
\Delta V ~=~- \widetilde{m}_\lambda \,Re \Big({\mu^3 \over 16}\, \phi_\chi
~+~ \gamma \,\phi^3 \Big), 
\label {deltaV}
\end{eqnarray}
where 
$\widetilde{m}_\lambda \equiv {32 \pi^2 \over g^2  N_c}\, m_\lambda$.

The expression (\ref {deltaV}) is the only correction to the 
effective potential to  leading order in $m_{\lambda}$. 
All higher order corrections are  suppressed by powers of 
$m_\lambda / \mu$. Those corrections are  neglected in this work.
\vspace{0.3in}\\
{\bf 4. The Mass Spectrum }
\vspace{0.2in}

Having derived the potential of the broken theory 
one turns to the calculation of the mass spectrum.
The potential consists of two parts, $V_0$ defined in (\ref
{Potential}) and the SUSY breaking term (\ref {deltaV}) 
\begin{equation}
V=V_0+\Delta V. \nonumber
\end{equation}
One calculates minima of  the full scalar potential $V$.
Explicit  though tedious  calculations yield the following results.
The VEV of the $\phi$ field does not get shifted
when the soft SUSY breaking terms are introduced.  Thus, even  in the 
broken theory $\langle \phi   \rangle =\mu $. However,  
the $\phi_\chi$  (and $B$ ) fields  acquire  nonzero VEVs 
in the broken case
\begin{eqnarray}
\langle \phi_\chi \rangle  = 
{8 \over 9 \alpha \mu} \widetilde{m}_\lambda ~~~~~{\rm and }~~~~~~
\langle B \rangle=-{8 \delta \over 9 \alpha }\widetilde{m}_\lambda \mu .
\label{VEV}
\end{eqnarray}
The shift of the vacuum energy  causes  the spectrum of the model to be 
also rearranged. Explicit calculations of the masses of all 
lowest-spin states yield  the following results
\begin{equation}
\label{mscalar}
M^2_{scalar\,\pm}~ ~=~~ M^2_{\pm} ~-~ 
{3 \over 4}\, \alpha \gamma\, \mu \,\widetilde{m}_\lambda \Bigg (
1~ \pm ~ \sqrt {1+x}~ \Bigg ) \Bigg (2~\pm~ {1\over \sqrt {1+x}}~
\Bigg),
\end{equation}

\begin{equation}
\label{mfermion}
M^2_{fermion \,\pm} ~=~M^2_{\pm} ~-~
{3 \over 4}\, \alpha \gamma\, \mu \,\widetilde{m}_\lambda \Bigg(
1 \pm \sqrt {1+x}~ \Bigg ) \Bigg (3 ~\pm~ {1\over \sqrt {1+x}}~
\Bigg),
\end{equation}

\begin{equation}
\label{mpseudo}
M^2_{p-scalar \,\pm} ~=~M^2_{\pm} ~-~
{3 \over 4}\, \alpha \gamma\, \mu \,\widetilde{m}_\lambda \Bigg (
1 \pm \sqrt {1+x} ~ \Bigg ) \Bigg (4~\pm~ {1\over \sqrt {1+x}}~
\Bigg),
\end{equation}
where $M^2_{\pm}$ denote the masses in the theory with unbroken SUSY
\cite {us}
\begin{equation}
\label{msusy}
M^2_{\pm}  ~=~ {18 \over (16)^2}{\alpha \over \delta}\mu^2 ~+~ {81 \over 2}
(\alpha \gamma)^2 \mu^2 \Bigg[1 ~\pm~ \sqrt{1 ~+~ x}~\Bigg],
~~~{\rm and}~~~ x\equiv {1 \over 288}
{\alpha \over \delta} {1\over(\alpha \gamma)^2}.
\nonumber
\end{equation}
In these expressions  the plus sign refers to the heavier
supermultiplet and the minus sign to the lighter set of states
\footnote[7]{In ref. \cite {us} we  used slightly different
notation. Masses in the heavy supermultiplet were denoted by $m_H$
and in the light supermultiplet by  $m_L$, so $M^2_\pm \equiv m^2_{H,L}$}.
One can verify that these values satisfy the mass sum rule to  
leading order in $O(m_{\lambda})$: 
\begin{eqnarray}
\label{sum}
\sum_{j} \,(-1)^{2j+1}\, (2j+1)\, M_j^2~=~ 0~,
\nonumber
\end{eqnarray}
where the summation goes over the spin $j$ of 
particles in the supermultiplet.

Let us discuss the mass shifts given in eqs. 
(\ref{mscalar}-\ref{mpseudo}). Consider 
the light  supermultiplet. In accordance with eqs. 
(\ref{mscalar}-\ref{mpseudo}), the masses in the light  multiplet are
increased  in the broken theory.  The biggest  mass shift is 
found in the pseudoscalar channel. The smallest shift  is observed 
in the scalar channel. The fermion mass falls in  between these two
meson states. Thus, the lightest state in the 
spectrum of the model is the particle 
which without mixing would have been the scalar glueball.
There is a fermion state above that scalar. Finally, the
pseudoscalar glueball is heavier than those two states. 

Let us now turn to the heavy supermultiplet. In the broken 
theory the masses in  that multiplet get pulled down. However,
all states of the heavy multiplet are still heavier than any state
of the light multiplet in the domain of validity of our approximations.
The ordering of the states in the heavy supermultiplet is just the
opposite as in the light supermultiplet: the lightest state is the
pseudoscalar meson, the heaviest is the scalar, and the fermion, as
required, falls between them.  
The qualitative features of the  spectrum are shown in fig. 1. 
\vspace{0cm}
\epsfysize=24.0 cm
\begin{figure}[htb]
\epsfbox{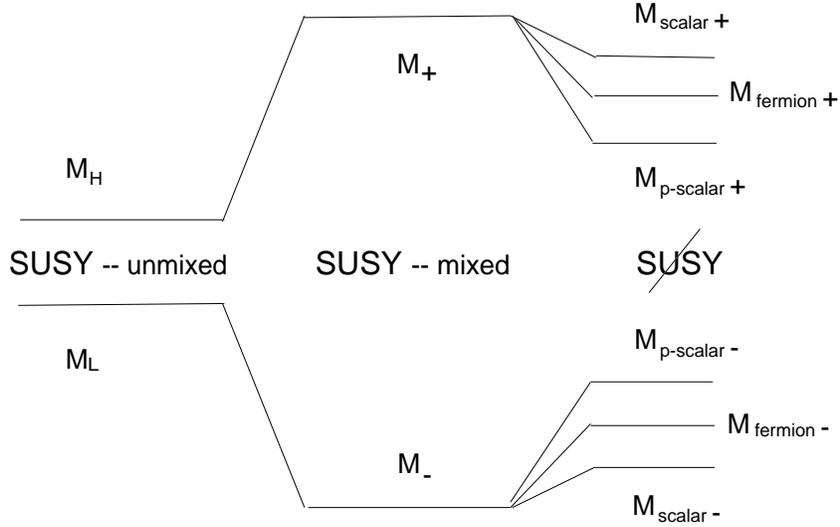}
\vspace{-16.5cm}
\caption{ Qualitative behavior of mass spectrum when passing from SYM to
softly broken model.
}
\label{phasediag}
\end{figure}

It is not surprising that the lowest mass state  obtained in
(\ref{mscalar}-\ref{mpseudo}) is a scalar particle. 
This is in  agreement with the result  of
ref. \cite{West} where  it was shown that the mass of the lightest
state which couples  to the operator $G_{\mu\nu}^2$ is less than the
mass of the  lightest state that couples  to $G \tilde{G}$, in pure
Yang-Mills theory. As a result, the lightest glueball turns out to be
the scalar glueball \cite{West}.  One can  apply the method  of ref.
\cite{West} to the SYM theory as well.  Due to the positivity of the
gluino determinant (see ref. \cite{steve}) one also deduces  that the
lightest state in softly broken SYM spectrum should be a scalar
particle. The pseudoscalar of that multiplet is therefore heavier.

Our result that the multiplet containing glueballs is split in such a
way that the scalar is lighter than the pseudoscalar, and vice versa
for the multiplet containing gluino-gluino bound states, is consistent
with expectations from quark-model lore.  In ordinary mesons the
$l=1$ states are heavier than their $l=0$ counterparts and the $l=0$
gluino-gluino bound state is a pseudoscalar, while an $l=0$
gluon-gluon bound state is a scalar.  It is interesting that in SYM
with massless gluinos the $l=0$ and $l=1$ bound states are degenerate,
but when the gluino masses are turned on one recovers the expected
ordering seen in $q \bar q $ states.

In the $\delta \rightarrow \infty$ limit one recovers  the VY
effective action.  The spectrum of the softly broken VY Lagrangian was
studied in \cite{NSM}. In that  limit only the heavy multiplet of the
spectrum survives. It is interesting that in the limit  $\delta
\rightarrow \infty$,   the ratio   of the mass-shifts of the surviving
states in (\ref{mscalar}-\ref{mpseudo}) is  $5:4:3$, which   differs
from the prediction of ref.  \cite{NSM}.  The seeming discrepancy is
resolved because in the limit $\delta \rightarrow \infty$ 
the vacuum expectation value of the glueball field $B$ 
tends  to infinity. Thus,  perturbing states about
that vacuum is not a well defined procedure. 
The right way  to obtain  the  $\delta \rightarrow \infty$
limit would be to decouple the ``glueball'' modes first,  and
then minimize the potential. This leads to a  shift of the
VEV of the $\phi$ field in the broken theory (as in ref. \cite {NSM}). 
As a result, the mass shifts calculated within this new vacuum state
are in agreement with the values reported  in the second work of 
ref. \cite {NSM}.  We stress, however, that on physical grounds we do
not expect SYM to realize the $\delta \rightarrow \infty $ limit of the
general effective Lagrangian (5).
\vspace{0.3in}\\
{\bf 5. Summary and Discussion }
\vspace{0.2in}

We have shown that the generalized VY effective action
can be written in two different ways. In one case the fundamental
superfield upon which the action is constructed is  the real 
tensor superfield $U$. In another approach all degrees
of freedom of the model are described by two chiral superfields
$\chi$ and $S$. In both cases the spectrum consists 
of two multiplets which are not degenerate in masses
even when SUSY is unbroken. 
The spin-parity quantum numbers  of these multiplets are identical to 
those of certain  chiral supermultiplets. 

The physical mass eigenstates are not pure 
gluon-gluon, gluon-gluino or gluino-gluino
composites; rather, the physical particles are mixtures of them.
The multiplet which without mixing would have been  
the glueball multiplet is lighter. As a result, those states cannot be 
decoupled  from the effective Lagrangian. This means that
comparisons  of lattice results to  analytic predictions
based on the original VY action are not justified.  

We introduced a  soft SUSY breaking term in the 
Lagrangian of the $N=1$ SUSY Yang-Mills model. The spurion method was used
to calculate the corresponding soft SUSY breaking terms in the generalized
VY Lagrangian. These soft breaking terms cause a shift of the vacuum energy 
of the model. The physical eigenstates which are  degenerate in the SUSY
limit, are  split when  SUSY breaking is introduced.  
We studied these mass splittings in detail. 
We have confirmed  that the spectrum of the broken theory
is in agreement with some low energy theorems \cite{West}, namely  
the scalar glueball turns out to be lighter than the pseudoscalar one.
The results of the present paper   can be directly tested in lattice
studies  of $N=1$ supersymmetric Yang-Mills theory (\cite {Lattice}).

\end{document}